\documentclass[runningheads]{llncs}

\usepackage[T1]{fontenc}

\usepackage[english]{babel}
\usepackage{subcaption}
\usepackage{url}
\usepackage{colortbl}
\usepackage{amssymb}
\usepackage{stmaryrd}
\usepackage{amsmath}
\usepackage{mathrsfs}
\usepackage{amssymb}
\usepackage[left]{lineno}
\usepackage{xfrac}
\usepackage{nicefrac}
\usepackage[textsize=scriptsize,backgroundcolor=yellow!40,obeyFinal]{todonotes}
\usepackage[nonumberlist,acronym,sanitize=none]{glossaries}
\glsdisablehyper
\usepackage{comment}
\usepackage[pdftex, colorlinks=true, hyperfootnotes=true, hyperindex=true,
             plainpages=false, pagebackref=false, pdfpagelabels=true, pdfstartview=FitH,
             linkcolor=purple, citecolor=teal, urlcolor=blue,
             bookmarks, bookmarksopen, bookmarksdepth=3]{hyperref}
\usepackage[capitalise,nameinlink]{cleveref}
\crefname{algocf}{alg.}{algs.}
\Crefname{algocf}{Algorithm}{Algorithms}
\crefname{section}{Sect.}{Sects.}
\Crefname{section}{Section}{Sections}
\usepackage{tkz-base}
\usetikzlibrary{decorations.pathmorphing,trees,snakes,arrows,shapes,automata,petri}
\usepackage{paralist}
\usepackage{multicol}
\usepackage{booktabs}
\usepackage{enumitem}
\usepackage{multirow}
\usepackage{rotating}
\usepackage{etaremune}
\usepackage[ruled,linesnumbered,algo2e]{algorithm2e}
\usepackage{marginnote}
\usepackage{mathtools}
\usepackage{adjustbox}
\usepackage{ifthen}
\usepackage[normalem]{ulem}
\usepackage{lineno}
\usepackage{soul}
\usepackage{floatrow}
\usepackage{listings}
\crefname{lstlisting}{Listing}{Listings}
\usepackage{lipsum}
\usepackage{makecell}
\usepackage{diagbox}
\usepackage[scientific-notation=false]{siunitx}
\usepackage{footmisc}
\usepackage{xspace}
\usepackage{ifdraft}
\usepackage{wrapfig}
\usepackage{orcidlink}
\newacronym[\glslongpluralkey={Distributed Ledger Technologies}]{dlt}{DLT}{Distributed Ledger Technology}

\newacronym{ipfs}{IPFS}{InterPlanetary File System}
\newcommand{\IPFS}[0] {\Gls{ipfs}\xspace}

\newacronym{p2p}{P2P}{peer-to-peer}

\newacronym{abe}{ABE}{Attribute-Based Encryption}
\newcommand{\ABE}[0] {\Gls{abe}\xspace}

\newacronym{maabe}{MA-ABE}{Multi-Authority Attribute-Based Encryption}
\newcommand{\MAABE}[0] {\Gls{maabe}\xspace}

\newacronym{cpabe}{CP-ABE}{Ciphertext-Policy Attribute-Based Encryption}

\newacronym{ssl}{SSL}{Secure Sockets Layer}

\newacronym{dht}{DHT}{Distributed Hash Table}

\newacronym{mpc}{MPC}{Multi-party Computation}

\newglossaryentry{box}{ %
  name={Box},
  description={authority intialisation storage box},
  first={\glsentrydesc{box} (henceforth, \glsentrytext{Box} for short)},
  plural={Boxes},
  firstplural={\glsentrydesc{box}es (henceforth, \glsentryplural{box} for short)}
}

\newglossaryentry{rloc}{ %
	name={resource locator},
	description={a generic term to denote content-based links obtained via hashing (e.g., the IPFS link)},
}

\newacronym{cpmaabe}{CP-MA-ABE}{Ciphertext-Policy Multi-Authority Attribute-Based Encryption}

\newglossaryentry{downer}{ %
  name={data owner},
  description={provide data for decision making},
  plural={data owners},
}

\def\DOwner{\glsentrytitlecase{downer}{name}\xspace}
\def\AttCert{Attribute Certifier\xspace}
\def\Reader{Reader\xspace}

\def\Auth{Authority\xspace}
\def\Auths{Authorities\xspace}

\newacronym{dk}{\textsl{dk}}{decryption key}
\newacronym{fdk}{\textsl{fdk}}{final decryption key}

\newacronym[\glslongpluralkey={Business Processes}]{bp}{BP}{Business Process}
\newacronym{bpi}{BPI}{Business Process Intelligence}
\newacronym{bpm}{BPM}{Business Process Management}
\newacronym{bpms}{BPMS}{Business Process Management System}
\newacronym{bpmn}{BPMN}{Business Process Model and Notation}
\newacronym{cpn}{CPN}{colored Petri net}
\newacronym{kpi}{KPI}{Key Performance Indicator}
\newacronym{ocbc}{OCBC}{Object-centric Behavioral Constraints}
\newacronym{soa}{SOA}{Service-Oriented Architecture}
\newacronym{pn}{PN}{Petri net}
\newacronym{wf}{WF}{workflow}
\newacronym{wfms}{WfMS}{Workflow Management System}
\newacronym{xes}{XES}{eXtensible Event Stream}
\newacronym{yawl}{YAWL}{Yet Another Workflow Language}
\newglossaryentry{task}{%
	name={task},description={the non-divisible, elementary activity}}

\newglossaryentry{promod}{%
	name={process model},description={the model of a process}
}
\def\LogAlph {\ensuremath{\Sigma}}
\newglossaryentry{logalph}{
	name={log alphabet},description={the process alphabet, as reflected in a log},%
	symbol={\LogAlph}}
\def\Evt {\ensuremath{e}}
\newglossaryentry{evt}{
	name={event},description={a record of an instantaneous fact during the process enactment},%
	symbol={\Evt}}
\def\Trc { \ensuremath{\tau} }

\newglossaryentry{trace}{
	name={trace},description={a sequence of \glsplural{evt}},%
	symbol={\Trc}}
\def\EvtLog {\ensuremath{L}}

\newglossaryentry{evtlog}{
	name={event log},description={a collection of \glstext{evttrace}s},%
	symbol={\EvtLog}}

\renewcommand\tabcolsep{5pt}
\newcolumntype{d}{>{\columncolor{gray!10}}c}
\newcolumntype{m}{>{\columncolor{gray!10}}l}
\newfloatcommand{capbtabbox}{table}[][\FBwidth]
\newenvironment{iiilist}%
{\begin{inparaenum}[\itshape(i)\upshape]}%
{\end{inparaenum}}

\RequirePackage{xparse}
\NewDocumentEnvironment{AuthNote}{+o+o}{%
	\IfValueT{#2}{\marginnote{\scriptsize{#2}}}%
	\begin{scriptsize}
		\colorbox{gray}%
		{\color{white} Note\IfValueT{#1}{ (#1)}:}%
		\quad%
		\color{brown}
}{%
	\normalcolor
	\end{scriptsize}
}

\RequirePackage{pifont}

\RequirePackage{lipsum}
\newcommand{\LipsumGray}[1][]{{\color{gray}\ifthenelse{\equal{#1}{}}{\lipsum}{\lipsum[#1]}}}

\RequirePackage{siunitx}
\newcolumntype{D}[1]{S[
	table-omit-exponent,
	round-mode=places,
	round-integer-to-decimal,
	round-precision={#1}]} 

\providecommand{\eg}{{e.g.,}\xspace}

\providecommand{\confetty}{{CONFETTY}\xspace}
\providecommand{\processInt}{{Process Interface}\xspace}
\providecommand{\confidInt}{{Confidentiality Interface}\xspace}

\newcommand{\SmallCode}[1]{\footnotesize\texttt{#1}\normalsize}

\graphicspath{{./images/}}

\newenvironment{Acknwoledgments}{\smallskip\par\noindent\footnotesize\textbf{\ackname}}{\normalsize}

\begin{document}

\title{CONFETTY: A Tool for Enforcement and Data Confidentiality on Blockchain-Based Processes}

\titlerunning{CONFETTY}

\author{
	Michele~Kryston\inst{1}\orcidlink{0009-0000-1491-2471}
	\and
	Edoardo~Marangone\inst{2}\orcidlink{0000-0002-0565-9168}
	\and \\
	Alessandro~Marcelletti\inst{3}\orcidlink{0000-0003-1192-6696}
	\and
    Claudio~{Di~Ciccio}\inst{1}\orcidlink{0000-0001-5570-0475}
}
\authorrunning{M. Kryston et al.}

\institute{
	Utrecht University, Utrecht, the Netherlands,
	\email{\href{mailto:m.kryston@uu.nl}{m.kryston@uu.nl}}; 
	\email{\href{mailto:c.diciccio@uu.nl}{c.diciccio@uu.nl}}
	\and
    Sapienza University of Rome, Rome, Italy
	\email{\href{mailto:edoardo.marangone@uniroma1.it}{edoardo.marangone@uniroma1.it}}
	\and
	University of Camerino, Camerino, Italy,
	\email{\href{mailto:alessand.marcelletti@unicam.it}{alessand.marcelletti@unicam.it}}   
}

\maketitle
\begin{abstract}
Blockchain technology enforces the security, robustness, and traceability of operations of Process-Aware Information Systems (PAISs). In particular, transparency ensures that all data is publicly available, fostering trust among participants in the system. Although this is a crucial property to enable notarization and auditing, it hinders the adoption of blockchain in scenarios where confidentiality is required, as sensitive data is handled. Current solutions rely on cryptographic techniques or consortium blockchains, hindering the enforcement capabilities of smart contracts and the public verifiability of transactions. This work presents the CONFETTY open-source web application, a platform for public-blockchain based process execution that preserves data confidentiality and operational transparency. We use smart contracts to enact, enforce, and store public interactions, while we adopt attribute-based encryption techniques for fine-grained access to confidential information. This approach effectively balances the transparency inherent in public blockchains with the enforcement of the business logic. 
\end{abstract}

\section{Introduction}
\label{sec:introduction}
Public permissionless blockchains
enable new forms of trustworthy Process-Aware Information Systems (PAISs) and inter-organizational collaborations~\cite{BCopportunities}, thanks to their ensemble of characteristics.
Distributed ledgers and consensus protocols underpin a shared, tamper-proof repository where multiple parties can store data without a central authority. Cryptographic primitives provide security guarantees and support accountability for recorded actions. Smart contracts are programs deployed and executed within the blockchain environment, which encode and automatically enforce business logic. In this environment, transparency is crucial since both on-chain data and the operations performed by smart contracts are visible and verifiable by everyone.

Collaborative process execution and monitoring require properties such as business logic enforcement and transparency~\cite{DiCiccio.etal/SoSyM2022:BlockchainForProcessMonitoring}. However, in many blockchain-based approaches, exchanged data is publicly accessible, making their adoption impractical when confidentiality is essential. Therefore, finding an appropriate balance between transparency and confidentiality is fundamental in domains that handle sensitive information.
Existing solutions typically address this issue in two main ways. On the one hand, private permissioned blockchains are used to restrict data visibility. 
While effective at hiding data, this approach reduces transparency for external auditors and often relies on strong security assumptions about the consortium system. On the other hand, cryptographic techniques enable fine-grained access control, so that only authorized parties can read the data~\cite{Koepke.etal/FGCS2023:DesigningSecureBusiness,Marangone.etal/EDOC2023:MARTSIA,Marangone.etal/BPM2022:CAKE}. Despite their benefits, their integration with process management systems is limited or non-existent.

To overcome this issue, we present the tool implementation of CONFidentiality EnforcemenT TransparencY (\confetty)~\cite{CONFETTY}, an approach and an architecture for blockchain-based process enactment that preserves the confidentiality of exchanged information while keeping public enforcement and transparency of process execution. 
In particular, we encode the business process logic of the interactions between parties in smart contracts, and we rely on \acrfull{maabe}~\cite{MAABE} to provide fine-grained access control over activity payloads and information artifacts, ensuring that only authorized parties can access sensitive data.

The remainder of the paper is organized as follows. \Cref{sec:background} presents the technologies of the framework, while \cref{sec:illustrativeExample} introduces a running example. In \cref{sec:innovationsAndFeatures}, we present \confetty's features by depicting its application in the running example.
\Cref{sec:maturity} evaluates the maturity of the framework.
In \cref{sec:relatedWork}, we review related work. Finally, \cref{sec:conclusionAndFutureWork} concludes the paper and proposes directions for future research and development.

\section{Background}
\label{sec:background}
Before explaining \confetty, we discuss its conceptual pillars in this section.

A \textbf{blockchain} is a distributed append-only ledger that records transactions, namely asset transfers among accounts, in a tamper-proof manner. Transactions are organized into blocks that contain a set of entries (among other things) and a cryptographic hash that links each block to its predecessor, forming a chain. 
Most blockchains (\eg Ethereum
) support the execution of smart contracts
, namely 
programs executed in a decentralized manner via virtual machines (\eg the Ethereum Virtual Machine, EVM)~\cite{Wood/2014:Ethereum}. 

Performing a transaction incurs a cost on the invoker. This cost depends on the complexity of the on-chain operation and the size of the exchanged data. 
Therefore, users may employ additional technologies, such as distributed hash-table tamper-proof storage systems, to deal with large amounts of data. In this way, only a low-cost unique link to that file is saved on-chain. An example of such technology is \textbf{\IPFS}~\cite{IPFS}, a peer-to-peer network for decentralized data storage that provides access to resources via a unique content-addressed, tamper-proof locator.  

\textbf{\ABE} is a public-key encryption scheme that links encrypted data with decryption keys via attributes
. Such a scheme can be used to fine-grain access to data stored on \IPFS for specific users. 
Typically, one authority generates decryption keys, introducing a single point of failure. 
\MAABE~\cite{MAABE} is a variant of \ABE that removes this issue by improving decentralization. 
In the Ciphertext-Policy version of MA-ABE (MA-CPABE)~\cite{MultiAuthorityCP}, users are associated with attributes, and data is encrypted using logical formulas (policies) created upon such attributes.

\section{Running Example}
\label{sec:illustrativeExample}
\begin{figure}[tb]
	\centering
	\includegraphics[width=\linewidth]{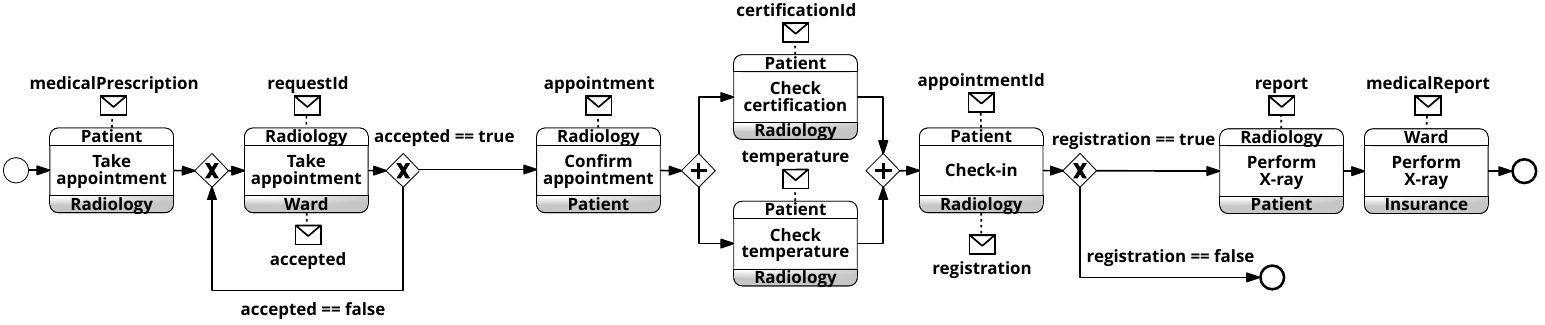}
	\caption{BPMN choreography diagram of an X-ray diagnostic analysis~\cite{runningExample}}
	\label{fig:running_example}
\end{figure}

\Cref{fig:running_example} illustrates an example in the healthcare domain inspired by~\cite{runningExample} that we will use throughout the paper to provide a clear description of \confetty. 
First, the patient 
presents the medical prescription and asks for an appointment. Then, the radiology department checks the ward's availability, and the appointment is confirmed if a date is available. Otherwise, a new tentative date is proposed. Once the patient's registration details are collected, the patient's temperature and vaccination certificate are verified. A radiology clerk subsequently evaluates whether the appointment can be confirmed. If so, the X-ray exam is performed, and the results, along with a final report, are delivered to the patient. Finally, an insurance agency receives the patient's results and may access the medical report to activate compensatory actions under the subscribed contract (\eg the reimbursement of medical bills). In addition to the participants shown in the choreography of~\cref{fig:running_example}, we assume the presence of the local Ministry of Health's inspector. Although not directly involved in the execution, the inspector must retain full access to the exchanged information for auditing purposes even after the process instance completes.

\section{Innovation and Features}
\label{sec:innovationsAndFeatures}
\begin{figure}[t]
	\centering
    	\includegraphics[width=\textwidth]{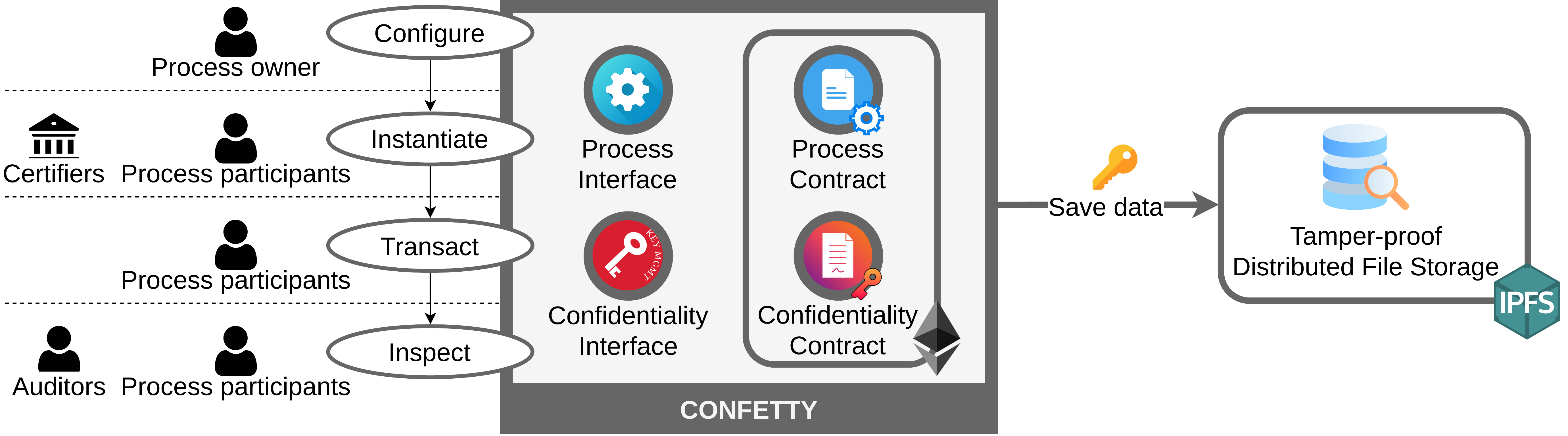}
	\caption[An overview of the \confetty architecture and functionalities]{An overview of the \confetty architecture and functionalities}
	\label{fig:confetty:architecture}
\end{figure}

\confetty is the first implemented platform enabling public-blockchain based process execution with fine-grained access control mechanisms to safeguard confidential data exchanges. 
\Cref{fig:confetty:architecture} diagrammatically depicts its architecture and functionalities. In this section, we explain them and illustrate the implemented prototype at work with the example described in~\cref{sec:illustrativeExample}.

\confetty relies on two external infrastructures used as architectural buttresses:
\begin{iiilist}
	\item A \textbf{{Programmable Blockchain Platform}} (in our implementation, Ethereum),
	\item and a \textbf{{Tamper-proof Distributed File Storage}} (here, IPFS). 
\end{iiilist}
The former maintains the public state storing transactions and runs smart contracts that implement and execute the process logic.
The latter (which we will henceforth refer to as Distributed FS for brevity) stores 
 the encrypted data. 

\confetty is comprised of four main components.
\begin{iiilist}
	\item The \textbf{{\processInt}} handles all the participants' operations that are related to the process. It acts as an interface to the blockchain, abstracting from the technical implementation and exposing functionality to users.
	\item The \textbf{{Process Contract}} is a smart contract that manages the process instances and their execution. It elaborates and manages all process specifications and their public states. 
	\item The \textbf{{\confidInt}}, built upon \MAABE, is responsible for handling participants' authorizations and operations involving confidential data. 
	It operates as a fa\c{c}ade towards the blockchain to store access grants, distributes decryption keys to the users, and handles the storage and retrieval of encrypted data.
	\item The \textbf{{Confidentiality Contract}} is a smart contract that notarizes the authorizations over confidential data, and their writing and access operations.
\end{iiilist}

\confetty provides four main functionalities.
\Cref{fig:instantiation} shows the graphical user interface designed to support the first two: \emph{Configure} and \emph{Instantiate}.
\begin{iiilist}
	\item The {\textbf{Configure}} 
    functionality allows process owners to specify the process to run on-chain. 
	The process owner sends the process specification files to the \processInt, which elaborates the inputs and stores the behavioral and business logic on-chain. Then, the \processInt composes the access policies for the confidential data. The process owner can revise these policies by adding custom roles that are not expected active parties in the process but need access to data (\eg the Ministry of Health). 
	Once policies are confirmed, they are sent to the \confidInt, which stores them in the Distributed FS and saves the generated locator on-chain.
\begin{figure}[tb]
    \centering
    \includegraphics[width=1\linewidth]{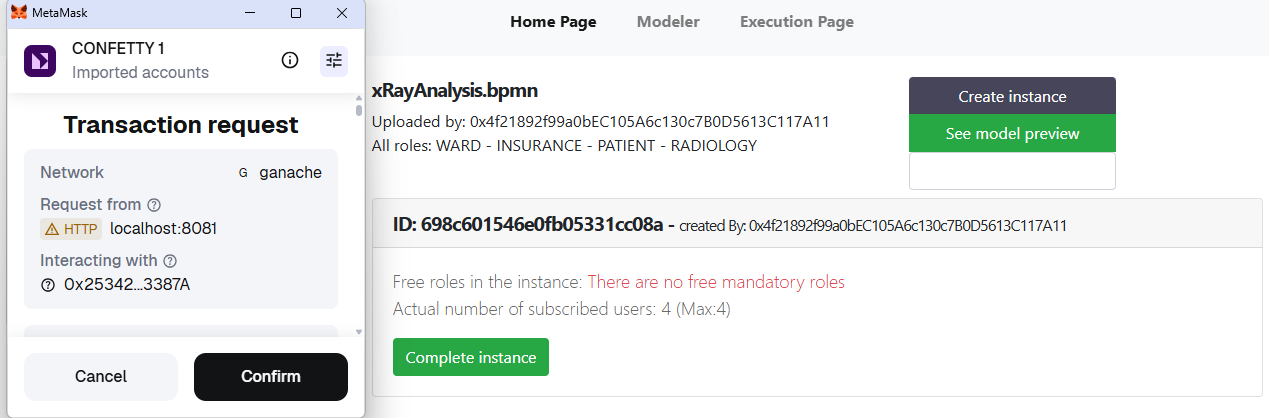}
    \caption{Kick-start of the process instance setting its public state on-chain}
    \label{fig:instantiation}
\end{figure}
%
\item The {\textbf{Instantiate}} functionality allows users to authenticate themselves as participants in the process, specifying their role via the \processInt.
Certifiers attest to authentication and confirm the user's self-declared roles. The \processInt merges the newly received information with the process data acquired in the Configure functionality. Thereupon, a selected participant kick-starts a new process instance via the \processInt, thereby initializing the instance's public state by sending a transaction. 

\Cref{fig:send_read} shows \confetty's graphical user interface for the \emph{Transact} and \emph{Inspect} functionalities.
\item The {\textbf{Transact}} functionality allows process participants to exchange information at runtime, 
in the form of both public and confidential data. 
Public data is directly stored on-chain, becoming publicly available and usable for control-flow decisions. 
A user sends execution data to the \processInt, which invokes the Process Contract to request a state update and forwards the execution data. 
This step also advances the process, updating the expected next elements and participants. 
To handle confidential information, the {\processInt} receives and forwards the data to the \confidInt, which retrieves the policy from the Distributed FS and uses it to encrypt the data via \MAABE, ensuring that only authorized participants can access it.
Then, the \confidInt stores the encrypted data in the Distributed FS, its locator in the Confidentiality Contract, and notarizes the operation on-chain.
\begin{figure}[tb]
    \centering
    \includegraphics[width=\linewidth]{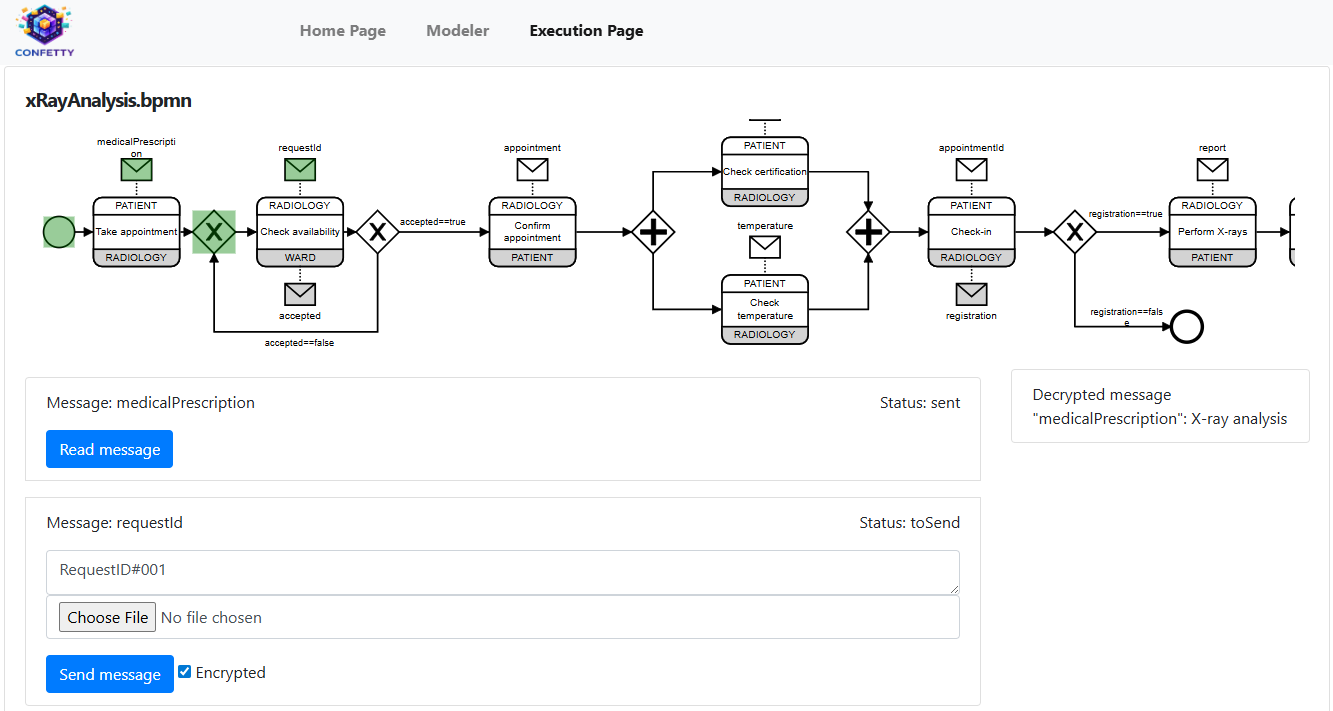}
    \caption{\confetty interface showing examples of message send and read}
    \label{fig:send_read}
\end{figure}
%
\item The {\textbf{Inspect}} functionality allows participants and auditors to access data. Public data can be freely retrieved, while confidential data is restricted to authorized users only. 
In the latter case, the user requests a decryption key via the \confidInt. Then, the \confidInt forwards the request to the Confidentiality Contract, which notarizes it on the Distributed Ledger for traceability.
The \confidInt generates an attribute-based key (or \textit{a-b key} for short) for the user who requested it. 
The a-b key, as the name suggests, encodes the user's attribute as per the \ABE scheme.
Once the user receives the a-b key, they can invoke the \confidInt to read the confidential data.
Then, the \confidInt retrieves the encrypted data from the Distributed FS and forwards it to the user. 
Only if the user's attributes satisfy the policy specified during data encryption can the user's a-b key decrypt the confidential data.
\end{iiilist}

\paragraph{A note on decryption keys.} We remark that an a-b key is associated with the individual user. The same key will be used to access all messages. In our example, consider a user who holds the attributes characterizing them as a \SmallCode{WARD} in the context of the case \SmallCode{PID476948}.
Their a-b key can decrypt the requestId for that case, but not the appointment.
The a-b key of a \SmallCode{PATIENT} user participating in the same case (\SmallCode{PID476948}) can decrypt the appointment instead, but not the requestId.
Notice that an \SmallCode{INSPECTOR} from the Ministry of Health, instead, can use their unique a-b key to access all the aforementioned documents regardless of the related process instance.
Since their individual key is the sole instrument for users to access a document, the \confidInt does not need to generate a different decryption key for every document and give the same copy to multiple users (which would be necessary with symmetric encryption~\cite{DBLP:journals/toit/LinHZHL21}), nor create numerous copies of the same document for different decryption keys (one per user, if an asymmetric encryption scheme was employed~\cite{Koepke.etal/FGCS2023:DesigningSecureBusiness}). Therefore, our approach saves unnecessary data replications and key distributions, while ensuring integrity. Also, as the key stays with the users, and the documents are stored and notarized on tamper-proof external infrastructures (the Distributed FS and the Programmable Blockchain Platform, respectively), the information remains available even in case the \confetty system stops functioning.

\section{Maturity}
\label{sec:maturity}
An Apache Tomcat 9 application server%
\footnote{Apache Tomcat:~\href{https://tomcat.apache.org/}{\nolinkurl{tomcat.apache.org}}; go-ipfs (now Kubo):~\href{https://github.com/ipfs/kubo}{github.com/ipfs/kubo}; Sepolia:~\href{https://sepolia.dev/}{\nolinkurl{sepolia.dev}}; Ganache:~\href{https://archive.trufflesuite.com/ganache/}{\nolinkurl{archive.trufflesuite.com/ganache}}; Web3.js: \href{https://web3js.readthedocs.io/en/v1.10.0/}{\nolinkurl{web3js.readthedocs.io}}. Accessed: \today. \label{foot:technologies}}
hosts the Process Interface and the Confidentiality Interface, implemented with JDK 13 and Python 3.6.9, respectively. Users interact with the server via the TLS protocol.
Go-ipfs v.~0.7
operates as the Distributed Tamper-proof File Storage.%
\textsuperscript{\ref{foot:technologies}}
Sepolia and Ganache%
\textsuperscript{\ref{foot:technologies}}
serve as Programmable Blockchain Platforms, the former as a test environment and the latter as a local RPC. Both are connected to the Process and Confidentiality Interfaces through Web3.js v.\ 1.10.0.\textsuperscript{\ref{foot:technologies}} Smart contracts are executed on the Ethereum Virtual Machine (EVM) and are written in Solidity v.\ 0.8.17.

We developed an initial prototype of \confetty to experimentally evaluate the performance of our approach~\cite{CONFETTY}, built upon the workflow management engine of ChorChain~\cite{corradini2021chorchain} and the data encryption/decryption mechanisms of MARTSIA~\cite{kryston2025martsia}. 
Compared to that preliminary release, this version of the \confetty tool comes packaged in a Docker container to enable its deployment and execution on a local machine. Also, it offers a refined graphical interface for users to interact with its functionalities. 
Results from literature-based processes and synthetic tests (varying the number of process participants, the choreography model size, the number of gateways, and the message payload size) show that off-chain execution time grows linearly and remains orders of magnitude smaller than the average Ethereum block time. Therefore, the tool imposes little to no overhead on the bandwidth for infrastructural information.
Regarding costs, they are comparable to or lower than those of state-of-the-art tools, although the latter only permit users to send process messages in clear.
Experimental evidence is reported in the aforementioned paper~\cite{CONFETTY}, alongside a discussion of the framework's robustness to cyberattacks with a threat model analysis.

\begin{sloppypar}
The source code and Wiki, offering a step-by-step tutorial on system setup and operation, are publicly available at \href{https://github.com/Process-in-Chains/CONFETTY}{\nolinkurl{github.com/Process-in-Chains/CONFETTY}%
}.
To watch a video showcasing the tool, visit \href{https://youtu.be/lj529HyqsbQ}{\nolinkurl{youtube/lj529HyqsbQ}}.
\end{sloppypar}

\section{Related Work}
\label{sec:relatedWork}
Lately, several solutions have been developed for (1) blockchain-based process automation and (2) preserving data confidentiality using cryptographic techniques. 
%
To automate collaborative processes (1), approaches including Caterpillar~\cite{Lopez-Pintado.etal/SPE2019:Caterpillar}, Lorikeet~\cite{Tran.etal/BPMDemos2018:Lorikeet}, and the aforementioned ChorChain~\cite{Corradini.etal/ACMTMIS2022:EngineeringChoreographyBlockchain}, have shown that blockchain-based solutions can strengthen trust among participants in multi-party collaborations and support monitoring and auditing~\cite{stiehle_process_blockchain_review}. They advance the integration of blockchain technology with process management, leveraging the security and traceability guarantees offered by distributed ledgers. However, they mainly address the control-flow perspective and do not provide mechanisms for secure, fine-grained access control to share data stored on public platforms.
%
To cater for data confidentiality on blockchain (2), encryption techniques have been leveraged. 
B-Box~\cite{B-Box}, CAKE~\cite{Marangone.etal/BPM2022:CAKE}, LU-MAABE-OD~\cite{Linjian} and the aforesaid MARTSIA~\cite{Marangone.etal/EDOC2023:MARTSIA} combine 
\IPFS, \ABE, and blockchain technology to this end, similarly to our approach. 
Although they 
address secure access control on blockchains, these solutions lack integration with process management systems.
\confetty is the first solution addressing both the above challenges.

\section{Conclusion and Future Work}
\label{sec:conclusionAndFutureWork}
In this work, we introduced \confetty, a platform for public-blockchain based process execution that combines transparency and public enforcement of the workflows while preserving secrecy of confidential data flows.
Process data for enforcement and auditability purposes is maintained on-chain as a public state, whose evolution is handled by smart contracts implementing the process control-flow and data-flow logic. We guarantee data confidentiality through the cryptographic scheme of \MAABE. In \confetty, data is encrypted, and access is granted only to authorized parties based on user attributes used to define access policies, which are bound to users' decryption keys.
Future work includes a field study to evaluate usability and adoption with real-world stakeholders. Further directions are the full decentralization of both architecture and governance 
and the design of mechanisms for automated decision support via computation over encrypted data.

\begin{credits}
\begin{Acknwoledgments}
	This work was partly funded by projects Health-e-Data and BRIE, funded by the EU-NGEU under the Cyber~4.0 NRRP MIMIT programme, and ASGARD, funded by Sapienza University of Rome under grant RG123188B3F7414.
\end{Acknwoledgments}
\end{credits}

\bibliographystyle{splncs04}
\bibliography{bibliography}

\end{document}